\documentclass[aoas,MSNbibl,nameyear,dvips]{arximspdf}

\doi{10.1214/13-AOAS614R} 
\referstodoi{10.1214/12-AOAS614}
\volume{7}
\issue{4}
\pubyear{2013}
\firstpage{1895}
\lastpage{1897}


\begin{document}
\begin{frontmatter}

\title{Rejoinder of ``Estimating the historical and future
probabilities of large terrorist events'' by~Aaron~Clauset~and~Ryan~Woodard}
\runtitle{Rejoinder}

\begin{aug}
\author[a]{\fnms{Aaron} \snm{Clauset}\corref{}\ead[label=e1]{aaron.clauset@colorado.edu}}
\and
\author[b]{\fnms{Ryan} \snm{Woodard}\ead[label=e2]{rwoodard@ethz.ch}}
\runauthor{A. Clauset and R. Woodard}
\affiliation{University of Colorado, Boulder and ETH Z\"urich}
\address[a]{Department of Computer Science and \\
\quad the BioFrontiers Institute\\
University of Colorado, Boulder\\
Boulder, Colorado 80309\\
USA\\
\printead{e1}}

\address[b]{Department of Management, Technology \\
\quad and Economics\\
ETH Z\"urich\\
Kreuzplatz 5, CH-8032 Z\"urich \\
Switzerland\\
\printead{e2}}
\pdftitle{Rejoinder of ``Estimating the historical and future
probabilities of large terrorist events'' by Aaron Clauset and Ryan Woodard}
\end{aug}

\received{\smonth{9} \syear{2013}}



\end{frontmatter}

We are pleased that our article has stimulated such thoughtful
discussion, and we appreciate the discussants' interest in exploring
the methodological and practical points of our study. The discussants
make a number of excellent points worthy of future investigation and
highlight the difficulty of making accurate statistical estimates of
the likelihood of rare events in general, and of large terrorist events
in particular. Our study is certainly not the final word on these
topics and we look forward to future developments in these areas. In
our rejoinder, we focus on selected points that will clarify the
context of our study and open questions, including the choice of
statistical models that are consistent with reasonable mechanistic
models for the data under study, and the value of simple models in
controlling uncertainty in complex social systems.

\textit{Not all tail models are equal}.
Two key motivations in our use of the power-law or simple Pareto tail
model were (i) its previous use in modeling terrorist event severities,
and (ii) its status as the only tail model with published mechanisms
for the frequencies of large terrorist events. Although the debate is
ongoing as to which mechanism, if any, is the correct explanation for
the observed heavy-tailed pattern in event severities [see~\citet
{clauset:gleditsch:2009} for discussion], these mechanisms provide an
important theoretical grounding for any statistical modeling of
terrorism's upper tail. \textit{Without such mechanisms, there is
little theoretical justification for favoring one particular tail model
over another to estimate extreme event probabilities.} Thus, we believe
some amount of priority should be given to estimates derived from
distributions like the power law, which have articulated and plausible
underlying mechanisms for terrorist event severities. However, the
surest way to reduce our ultimate uncertainty as to the likelihood of
future large events is to identify and test alternative mechanisms for
the heavy-tailed pattern in terrorist event severities, and we look
forward to new work in that direction.

\textit{Disagreement among tail models}.
The statistical framework we presented is entirely general and can thus
be used in conjunction with (i) any well-defined, automatic method for
identifying the upper tail region, and (ii) any well-defined
probabilistic model of an upper tail. (Although we modeled severities
as i.i.d. random variables, this is not a requirement, and a clear
understanding of the statistical correlations among terrorist events at
the global scale could, in principle, be incorporated into a more
detailed model.) Given these choices, data-driven estimates are then produced.

Even without regard to their theoretical motivation, not all tail
models are reasonable choices in this framework. Any model can be
fitted to the data, but if it is a poor fit, we are under no obligation
to trust its results. How then should we decide which models are good
fits? The models used in our analysis (power law, log-normal and
stretched exponential) were all previously demonstrated, under a
combination of standard hypothesis tests and likelihood ratio tests, to
meet this criteria~[\citet{clauset:etal:2009}]. Of course, more flexible
models, like the generalized Pareto distribution (GPD), the tapered
Pareto or a piecewise Pareto, may also provide reasonable fits, as
several discussants suggested. The difficulty, however, is how to
interpret or reconcile models that produce conflicting estimates for
the likelihood of a large event, and how to choose among models with
different levels of flexibility.

Modern statistics does not offer clear answers to these questions
because of the role played by $x_{\min}$, the smallest value for which
the tail model holds. Unlike traditional model parameters, changing
$x_{\min}$ changes the sample size, which confounds changes in
statistical power with the usual bias-variance trade-off. The result is
an additional risk in overfitting, particularly with flexible models
like those suggested by the discussants, as larger values of $x_{\min}$
are considered. Simple models, like the power-law distribution, would
seem to offer some protection against this risk. (We note that this
problem of identifying reasonable tail models is ubiquitous in complex
social systems with heavy-tailed variables, including financial
markets [\citet{farmer:lillo:2004,fcir:2011}].) The method we used to
choose $x_{\min}$ has reasonable properties and performs well in
practice [\citet{clauset:etal:2009}]. New research on how best to choose
$x_{\min}$ for out-of-sample forecasts and how to rigorously compare
models estimated from differently sized samples would help resolve some
of these difficulties.

\textit{Conclusions}.
The statistical modeling and forecasting of terrorist event severity is
a relatively new endeavor that combines interesting methodological
problems and tricky forms of uncertainty in an application area with
arguably genuine practical benefit. There are many interesting and
important questions worthy of study, and the discussants have
identified a number of them. Better estimates and a deeper
understanding of the pattern of large terrorist events and the
mechanisms that produce them can inform our expectations (as with large
natural disasters) of how many such events will occur over a long time
horizon and how to appropriately anticipate or respond to them.

In closing, we note that the relatively smooth distribution of the
sizes of terrorist events worldwide presents a puzzle. Given the highly
contingent nature of individual events and individual conflicts, 9/11
being an outstanding example, how can the global distribution be so
regular? This striking pattern suggests both that accurate estimates of
the probability of large events may be derived from modeling the
relative frequency of much less severe events and that some aspects of
global terrorism may not be as contingent or unpredictable as is often
assumed. Understanding the origin of this global-level pattern, and the
mechanisms by which local-level dynamics give rise to it, is itself an
important research direction with real implications for understanding
the fundamentals of violent political conflict. We look forward to new
insights in these directions.



%



\printaddresses

\end{document}